\documentclass[aps,prb,superscriptaddress,notitlepage,twocolumn]{revtex4-1}
\usepackage{hyperref}
\usepackage{amsmath}
\usepackage{graphicx}
\usepackage{color}
\usepackage{braket}
\usepackage{subfigure}
\usepackage{epstopdf}

\newcommand{\be}{\begin{equation}}
\newcommand{\ee}{\end{equation}}

\usepackage{bm}

\newcommand{\addcqt}{Centre for Quantum Technologies, National University of Singapore, 3 Science Drive 2, 117543, Singapore}
\newcommand{\addtuc}{School of Electrical and Computer Engineering, Technical University of Crete, Chania, 73100, Greece}
\newcommand{\addteda}{TEDA Applied Physics Institute and School of Physics, Nankai University, Tianjin, 300457, China}
\newcommand{\addsu}{Collaborative Innovation Center of Extreme Optics, Shanxi University, Taiyuan, Shanxi 030006, China}

\begin{document}

\title{Persistent homology analysis of a generalized Aubry-Andr\'{e}-Harper model}
% Force line breaks with \\

\author{Yu He}
\affiliation{\addteda}

\author{Shiqi Xia}
\affiliation{\addteda}

\author{Dimitris G. Angelakis}
\affiliation{\addcqt}
\affiliation{\addtuc}

\author{Daohong Song}
\affiliation{\addteda}
\affiliation{\addsu}

\author{Zhigang Chen}
\email[Email: ]{zgchen@nankai.edu.cn}
\affiliation{\addteda}
\affiliation{\addsu}

\author{Daniel Leykam}
\email[Email: ]{cqtdani@nus.edu.sg}
\affiliation{\addcqt}

\date{\today}

\begin{abstract}
Observing critical phases in lattice models is challenging due to the need to analyze the finite time or size scaling of observables. We study how the computational topology technique of persistent homology can be used to characterize phases of a generalized Aubry-Andr\'{e}-Harper model. The persistent entropy and mean squared lifetime of features obtained using persistent homology behave similarly to conventional measures (Shannon entropy and inverse participation ratio) and can distinguish localized, extended, and crticial phases. However, we find that the persistent entropy also clearly distinguishes ordered from disordered regimes of the model. The persistent homology approach can be applied to both the energy eigenstates and the wavepacket propagation dynamics.
\end{abstract}

\maketitle

\section{Introduction}

Wave propagation in low dimensional systems reveals many surprises thanks to wave coherence and interference effects~\cite{book,disorder_review,loc_review,fb_review,AL_experiment,Kohlert2019,Ganeshan2015,Wang2020,Goblot2020,An2021}. Disorder gives rise to Anderson localization and a complete suppression of transport; on the other hand, quasiperiodic potentials can exhibit localization transitions and critical phases supporting self-similar or fractal behaviour. Adding quantum or nonlinear interactions leads to even richer phenomena including subdiffusive spreading, many body localization, and quantum scars~\cite{nonlinear,MBL,scars}. While distinguishing localized from delocalized phases is relatively straightforward, the identification of critical and non-ergodic phases remains challenging, particularly when the relevant order parameters are unknown. Standard approaches involve studying the finite size or time scaling of simple observables over a huge range of scales~\cite{Anderson_review,Roemer2011}.

An alternative to finite size scaling is to identify relevant order parameters directly using machine learning~\cite{ML_review}, an in particular topological machine learning approaches are attracting growing interest as a means of identifying nonlocal order parameters~\cite{Ghrist2008,TDA_review,TDA_review2,Murugan2019,TDA_review3}. Specifically, the technique of persistent homology computes topological features over a range of scales, recording the scales at which features appear and are destroyed. This information is typically summarized using a persistence diagram, which forms a topological ``fingerprint'' of the data. Persistent homology has been applied to various physical systems, including the characterization of amorphous phases of matter and the detection of quantum phase transitions~\cite{Hiraoka2016,Mengoni2019,Olsthoorn2020,Tran2021,Leykam2021,Spitz2021,Cole2021,Tirelli2021,Sale2022,Park2022}.

Existing studies have largely used persistent homology as part of a larger machine learning pipeline, in which persistent homology identifies a large set of potentially-relevant order parameters, which are fed into clustering or dimensional reduction algorithms to learn the system's phase diagram. One drawback of this approach is that it is harder to understand than conventional methods, raising the question of whether persistent homology-based machine learning can be trusted to unveil new phenomena and not merely reproduce known phases. More cases studies are needed to better understand the strengths and limitations of persistent homology-based methods for identifying phase transitions.

In this article we apply persistent homology to study phase transitions in the Aubry-Andr\'{e}-Harper (AAH) model, a prototypical model of localization and criticality~\cite{AAH1,AAH2,Zilberberg,Ganeshan2013,Liu2015,Purkayastha2017,AAH_experiment}. We use persistent homology to obtain a compact encoding of the shape of the eigenfunctions across a range of wavefunction intensity scales in the form of persistence diagrams. To distinguish the different phases of the model we compute two summary statistics of the persistence diagrams: the root-mean-square feature lifetime and the persistent entropy. We find that while these summary statistics generally behave similarly to the conventional statistic measures (inverse participation ratio and Shannon entropy of the full eigenstate profiles), and they can be better at distinguishing the localized, extended, and critical phases, especially for the propagation dynamics.

Surprisingly, we find that persistent homology also distinguishes ordered from disordered states, revealing hidden order of the low energy eigenstates that is not detected by the inverse participation ratio or Shannon entropy. We show how this hidden order arises by performing a long wavelength expansion of the tight binding Hamiltonian, demonstrating that the quasiperiodic potential becomes transparent to low energy modes. Finally, we carry out numerical simulations of wavepacket propagation to show how the persistent homology approach can also distinguish the different phases using wavepacket dynamics.

The article is structured as follows: Sec.~\ref{sec:model} presents the phase diagram of the generalized Aubry-Andr\'{e}-Harper model and analyzes its low energy continuum limit. Sec.~\ref{sec:tda} briefly reviews persistent homology and shows how it can be applied to study the shape of eigenfunctions. Sec.~\ref{sec:dynamics} applies persistent homology to study wavepacket propagation dynamics. The article concludes with Sec.~\ref{sec:conclusion}.

\section{Generalized AAH model}
\label{sec:model}

\begin{figure}
    \centering
    \includegraphics[width=\columnwidth]{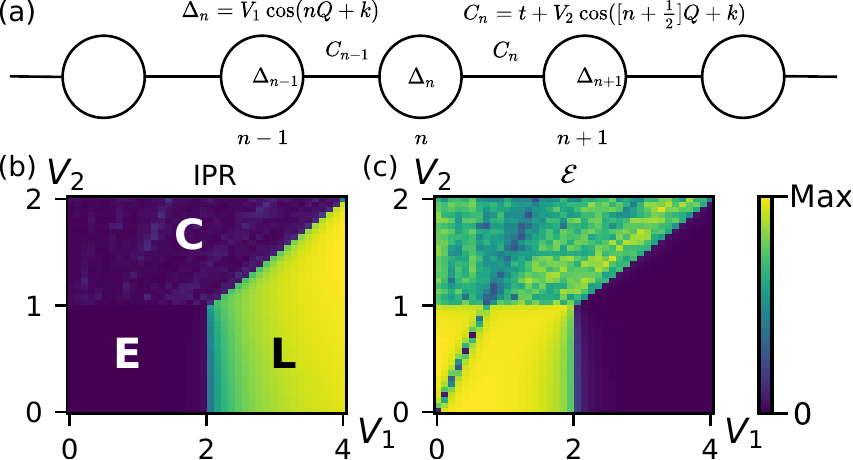}
    \caption{(a) Schematic of the generalized AAH model (Eq.~\eqref{eq:aah}) in which both the site energies and nearest neighbour couplings are quasiperiodically modulated, with modulation strengths $V_1$ and $V_2$ respectively. (b,c) Phase diagrams of the AAH model obtained using its ground state's inverse participation ratio IPR (b) and persistent entropy $\mathcal{E}$ (c), revealing extended (E), critical (C), and localized (L) phases. The dark straight line in (c) reveals the emergence of order in the low energy eigenstates, as explained in the text.}
    \label{fig1}
\end{figure}

We consider a generalized Aubry-Andr\'{e}-Harper model introduced in Ref.~\cite{Liu2015}, corresponding to a one-dimensional tight binding chain with quasiperiodically-modulated on-site energies and couplings illustrated in Fig.~\ref{fig1}(a). The lattice Hamiltonian is
\begin{align}
    \hat{H} \psi_n = V_1 \cos (nQ + k) \psi_n \nonumber \\
    +\{ t + V_2 \cos [(n+\frac{1}{2})Q + k]\} \psi_{n+1} \nonumber \\
    +\{ t + V_2 \cos [(n-\frac{1}{2})Q + k]\} \psi_{n-1}, \label{eq:aah}
\end{align}
where $V_1$ and $V_2$ are the on-site and coupling quasiperiodic modulation strengths, respectively, $t$ is the coupling in the absence of modulation, and the modulation has a frequency $Q$ and phase $k$. We consider the same parameters as in Ref.~\cite{Liu2015}: $Q=(1+\sqrt{5})\pi$, $t=1$, $0\leq V_1 \leq 4$, $0 \leq V_2 \leq 2$, $k=0$, and a lattice size of $N=300$ sites. We apply periodic boundary conditions to the wavefunction, $\psi_{n+N}=\psi_n$. Note that when $N$ and $Q$ are incommensurate, the periodic boundary conditions result in an effective defect in the modulation between sites $1$ and $N$. Similar to the original AAH model, there are no mobility edges and all eigenstates of Eq.~\eqref{eq:aah} are either localized, extended, or critical depending on the choice of parameters~\cite{Liu2015}.

First we will analyze the eigenstates of Eq.~\eqref{eq:aah} with energy $E$, $\hat{H} \phi_n = E \phi_n$, focusing on the properties of their probability density distribution $|\phi_n|^2$. Later, we will consider the propagation dynamics generated by Eq.~\eqref{eq:aah}, i.e. $i \partial_z \psi_n(z) = \hat{H}_n \psi_n(z)$, focusing on the case of light propagation in optical waveguide arrays, where one prepares an initial state $\psi_n(0)$ whose propagation along the longitudinal ($z$) direction is analogous to time evolution in the Schr\"odinger equation. We will apply a similar approach to analyze $z$-dependent intensity profiles $|\psi_n(z)|^2$; we will use probability density and intensity interchangeably in the following.

The localization of the normalized eigenstates ($\sum_n |\phi_n|^2 = 1$) is typically studied using the inverse participation ratio (IPR),
\be 
\mathrm{IPR} = \sum_n |\phi_n|^4,
\ee
which measures the reciprocal of the number of strongly excited sites~\cite{Anderson_review,Roemer2011}. In the large $N$ limit we have $\mathrm{IPR} \propto N^{-d}$, where $d$ is the fractal dimension of the eigenstate. $d = 0$ for localized eigenstates, e.g. exponential localization $|\phi_n|^2 \propto \exp(-|n|/\xi)$ yields $\mathrm{IPR} \sim 1/\xi$; $d=1$ for extended eigenstates, e.g. Gaussian random matrix ensemble eigenstates satisfy $|\phi_n|^2 = O(1/N)$, with $\mathrm{IPR} = 3/N$. On the other hand, critical eigenstates have $0 < d < 1$.

Another measure of an eigenstate's localization is the Shannon entropy of its probability distribution~\cite{entropy_0,entropy_1,entropy_2},
\be 
S = -\sum_n |\phi_n|^2 \ln |\phi_n|^2, \label{eq:shannon}
\ee 
which vanishes ($S=0$) when the probability density is concentrated at a single site, and attains its maximal value ($S=\ln N$) for delocalized states with modes distributed over all $N$ sites. For these simple profiles, one can obtain $S \sim -\ln (\mathrm{IPR})$.

Figure~\ref{fig1}(b) plots the inverse participation ratio of the ground state of $\hat{H}$ as a function of $V_1$ and $V_2$. $\mathrm{IPR}$ clearly distinguishes the extended and localized phases, but shows a poor contrast between the extended and critical phases, requiring finite size scaling to unambiguously distinguish them. This is because $\mathrm{IPR}$ is mainly sensitive to the number of strongly-excited sites, but misses information contained in the weak tails of the eigenstates. This can be addressed by considering other moments of the probability distribution, i.e. the generalized inverse participation ratios $\mathrm{IPR}_q = \sum_n |\phi_n|^{2q}$, corresponding to the $q$th moment of the probability distribution, to obtain a spectrum of fractal exponents $d_q$ (the multifractal spectrum~\cite{Anderson_review,Roemer2011}). We note that the Shannon entropy provides similar information to $\mathrm{IPR}_1$, and that the relation $S \sim -\ln (\mathrm{IPR})$ can be violated for eigenstates exhibiting multifractality.

Another limitation of the inverse participation ratio and Shannon entropy is that they are not strongly sensitive to the shape of the eigenstates. For example, both measures are invariant under permutations of the site positions and fail to distinguish a mode with a single, broad peak from one with multiple narrower peaks. While similarity measures for pairs of wavefunctions including the overlap correlation function~\cite{Cuevas2007} and Kullback-Liebler divergence~\cite{Luitz2015} are sensitive to changes in shape such as permutations in site positions, they cannot be applied to individual eigenstates and thus they provide a relative rather than absolute measure of shape. This motivates our investigation of whether persistent homology can provide a useful and more efficient characterization of the eigenstates.

Figure~\ref{fig1}(c) illustrates the phase diagram obtained using persistent homology analysis, which we will explain in the following section. In addition to clearly distinguishing the three phases, Fig.~\ref{fig1}(c) exhibits a local minimum in the eigenstate measure (persistent entropy $\mathcal{E}$) along the line $V_1 = 2V_2 \cos (Q/2)$, corresponding to the emergence of ordered eigenstates. This order can be understood by transforming to reciprocal ($p$) space, $\psi_n = \sum_p \psi_p e^{i p n}$, which yields (assuming $k=0$):
\begin{align}
\hat{H}\psi_p = &2 t \cos p  \; \psi_p \nonumber \\
&+ \left(\frac{V_1}{2} + V_2 [\cos p \cos \frac{Q}{2} - \sin p \sin \frac{Q}{2}]\right) \psi_{p+Q} \nonumber \\
&+ \left(\frac{V_1}{2} + V_2 [\cos p \cos \frac{Q}{2} + \sin p \sin \frac{Q}{2}]\right) \psi_{p-Q}.
\end{align}
When $V_1 = 2V_2 \cos (Q/2)$ the quasiperiodic potential becomes transparent to the low-energy modes, e.g. the plane wave $\phi_p = \delta_{p,\pi}$ forms an exact eigenstate (the ground state).

\begin{figure}
    \centering
    \includegraphics[width=0.6\columnwidth]{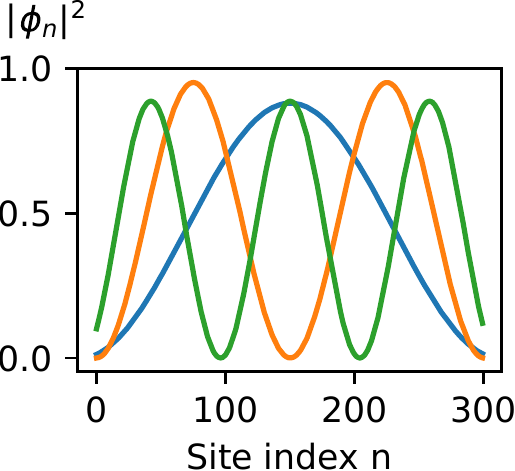}
    \caption{Probability density profiles (unnormalized) of the three lowest energy eigenstates on the ordered line shown in Fig.~\ref{fig1}(c) ($V_1 = 1 = 2 V_2 \cos (Q/2)$).}
    \label{fig:critical_line}
\end{figure}

To validate this intuition, Fig.~\ref{fig:critical_line} shows the three lowest energy eigenstates obtained numerically for $V_1 = 1 = 2V_2 \cos(Q/2)$, revealing simple standing wave forms with intensities $\sim \sin^2(n/N)$, $\sin^2(2n/N)$, $\sin^2(3n/N)$. The standing wave patterns can be understood by noting that the periodic boundary conditions $\psi_{n+N} = \psi_n$ are incommensurate with the quasiperiodic modulation, thus there is a (weak) effective defect potential in the vicinity of the $N$th lattice site; the low energy modes form standing waves that have a minimal overlap with this defect.

More generally, whenever some momentum $p$ satisfies $V_1 \leq -2 V_2 \cos (p \pm Q/2)$ there will be a decoupling of the momentum space eigenvalue problem into different sectors, which may explain the additional scar-like features seen in Figs.~\ref{fig1}(b,c).

In the extended phase the transparency of the quasiperiodic potential for low energy modes is missed by the standard eigenstate measures such as $\mathrm{IPR}$ and $S$, which only measure the localization of the modes and not their degree of disorder. This suggests that persistent homology may be sensitive to different features of the eigenstates, providing an additional tool for understanding complex lattice models. On the other hand, in the critical phase scars are visible in both $\mathrm{IPR}$ and $S$, since these measures can distinguish critical eigenstates from the extended states for which the quasiperiodic potential is transparent.

\section{Persistent homology of eigenfunctions}
\label{sec:tda}

Most previous applications of persistent homology to physics considered point cloud data equipped with a suitable metric, such as atomic positions and Euclidean distance for characterizing the structure of materials~\cite{Hiraoka2016}. This is a natural approach for dealing with sets of discrete objects. On the other hand, for analyzing individual eigenfunctions it is less straightforward to construct point clouds, requiring some way to reduce the continuous data to a discrete set of points, for example by choosing strong local maxima of the probability density distribution using some threshold~\cite{Spitz2021}.

An approach better-suited to probability density or intensity data is the sublevel set filtration, which characterizes the local maxima and minima of an image or intensity profile~\cite{survey}. Specifically, one considers the number of distinct low intensity clusters of the profile, i.e. the number of connected regions formed by sites with $|\phi_n|^2 < \epsilon$ as a function of $\epsilon$. As $\epsilon$ is continuously increased from zero, a new cluster will emerge whenever $\epsilon$ is tuned through a local minimum of the intensity. Neighbouring clusters will merge when $\epsilon$ crosses the local intensity maximum between them. Thus, the sublevel set filtration provides a stable way to characterize the critical points of an intensity or probability density profile. In effect, this allows one to count the number of distinct peaks in a profile over a range of intensity scales, with the lifetime (i.e. persistence) of each peak giving a measure of its significance. This is complementary to other approaches for studying localization in disordered systems such as fractal analysis, based on indirectly varying a characteristic intensity scale (associated with the eigenstate normalization) by changing the system size $N$~\cite{Anderson_review,Roemer2011}.

\begin{figure}
    \centering
    \includegraphics[width=\columnwidth]{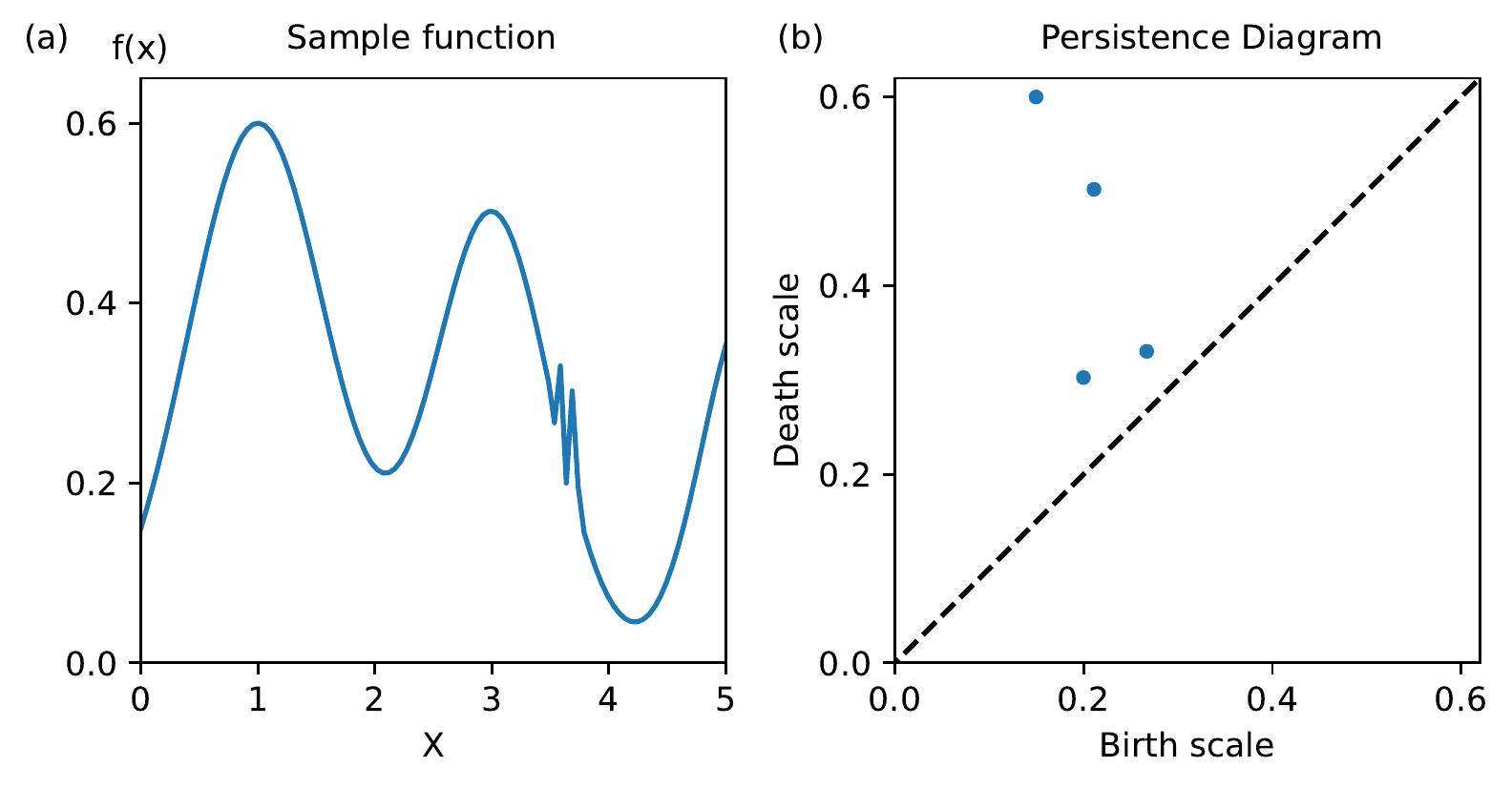}
    \caption{(a) A simple one-dimensional function $f(x)$ and its corresponding sublevel set persistence diagram (b), which summarizes its critical points (local maxima and minima).}
    \label{fig:PH_example}
\end{figure}

For example, Fig.~\ref{fig:PH_example} presents a simple intensity profile and its resulting persistence diagram. Each point in the persistence diagram represents a distinct local minimum of the intensity profile, with the horizontal (birth scale) and vertical (death scale) coordinates corresponding to the intensity at the local minimum and closest local maximum, respectively. All points occur above the diagonal (indicated by the dashed line), because the intensity at the local minimum will always be less than that at the nearest local maximum. The distance of a point to the diagonal, i.e. the intensity contrast between the local minima and maxima, provides a measure of the significance of the feature. Here there are two long-lived features (prominent peaks), and two short-lived features created by small fluctuations of the profile at $x \approx 3.6$. 

\begin{figure}
    \centering
    \includegraphics[width=\columnwidth]{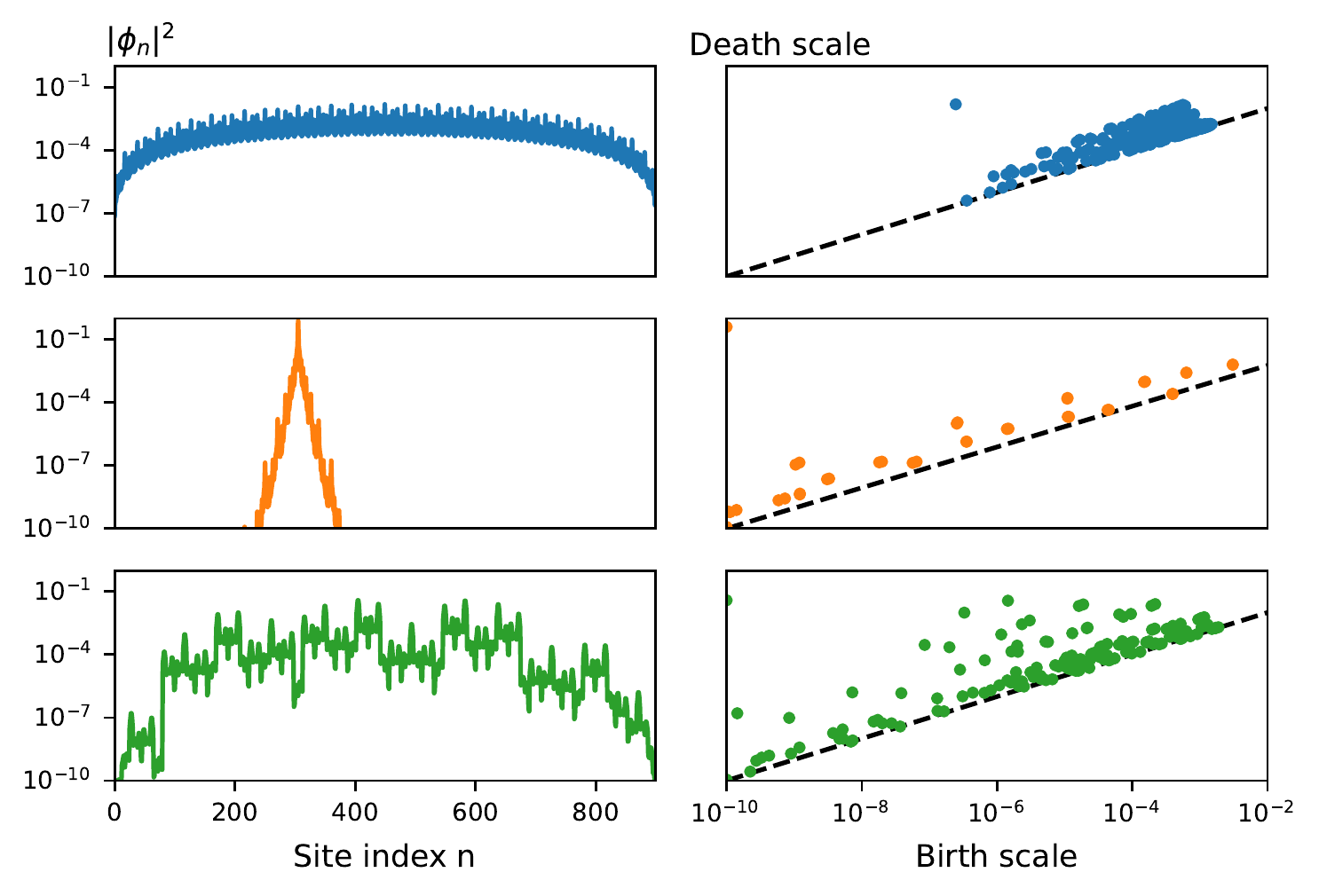}
    \caption{Probability density profiles (left) and feature lifetimes (right) of the lowest energy eigenstate of the AAH model in its three phases: Extended (top/blue; $V_1=1.8$, $V_2=0.5$), localized (middle/orange; $V_1=2.2$, $V_2 = 0.5$), critical (bottom/green; $V_1=1.0$, $V_2=1.5$).}
    \label{fig2}
\end{figure}

We now construct persistence diagrams for the ground states of Eq.~\eqref{eq:aah} in each of its three phases. Fig.~\ref{fig2} shows the eigenstates and corresponding persistence diagrams. In the extended phase the probability density of the ground state is oscillatory with significant intensity throughout the entire lattice, corresponding to all features (except one) being created and destroyed within a small range of intensities. 

With the increase of $V_1$ the generalized AAH model enters the localized phase, with the intensity localized to a very small interval, corresponding to only a single long-lived feature. Other features (induced by fluctuations in the eigenstate tails) have negligible lifetimes. 

In the critical phase the probability density distribution is not completely concentrated in a certain interval, with multiple disconnected local maxima. This results in many long-lived features with a range of birth scales, consistent with the self-similar (fractal) nature of the critical phase. 

The above sublevel set filtration approach of persistent homology summarizes all the critical points of an intensity profile, but cannot be directly used to quantify differences between the three phases. For example, how do we identify the transition from the extended and critical phases by inspecting the persistence of their features? One solution is to introduce a metric quantifying the distance between persistence diagrams, such as the persistence landscape~\cite{landscape}. An simpler alternative we will consider here is to compute summary statistics of the feature lifetime, reducing each persistence diagram to a single, more easily-interpretable number. 

We will consider two summary statistics. The first is the $p$-norm of the feature lifetimes~\cite{Tran2021}
\be 
\mathcal{P}_p(D) = \left(\sum_{(b,d)\in D} |d-b|^p \right)^{1/p}, \label{eq:pnorm}
\ee
where $b$ is the intensity at which a feature appears (i.e. a local minimum), and $d$ denotes the intensity at the local maximum at which the feature is destroyed. $d-b$ is thus the feature lifetime. For example, $\mathcal{P}_{\infty}$ measures the lifetime of the longest-lived feature in the persistence diagram $D$, $\mathcal{P}_1$ is the sum of feature lifetimes, and $\mathcal{P}_2$ is the root mean squared lifetime. Thus, the $p$-norms resemble the generalized inverse participation ratios.

The second summary is the persistent entropy~\cite{PH_entropy_stability},
\be 
\mathcal{E}(D) = -\sum_{(b,d)\in D} \frac{|d-b|}{\mathcal{S}(D)} \log \left( \frac{|d-b|}{\mathcal{S}(D)} \right), \label{eq:entropy}
\ee
where ${\mathcal{S}(D)} =  \sum_{(b,d)\in D}{|d-b|}$ is the sum of feature lifetimes in the persistence diagram D. The persistent entropy quantifies the non-uniformity of the feature lifetimes $d-b$; it is simply the Shannon entropy of the feature lifetime fractions.

Essentially, the standard approach for quantifying eigenstate localization is to compute summary statistics of their intensity distribution such as the inverse participation ratio or Shannon entropy. On the other hand, persistent homology computes and characterizes the eigenstates' critical points (local maxima and minima). In effect, we compress the information contained in the eigenstate profiles. The point summaries not only capture the localization properties of the eigenstate, but also its level of disorder. Referring back to Fig.~\ref{fig1}, we see that the persistent homology measures not only reproduce the known phase boundaries of the generalized AAH model, but also identify model parameters for which order is restored, i.e. along the line $V_1 = 2V_2 \cos (Q/2)$.

\begin{figure}
    \centering
    \includegraphics[width=\columnwidth]{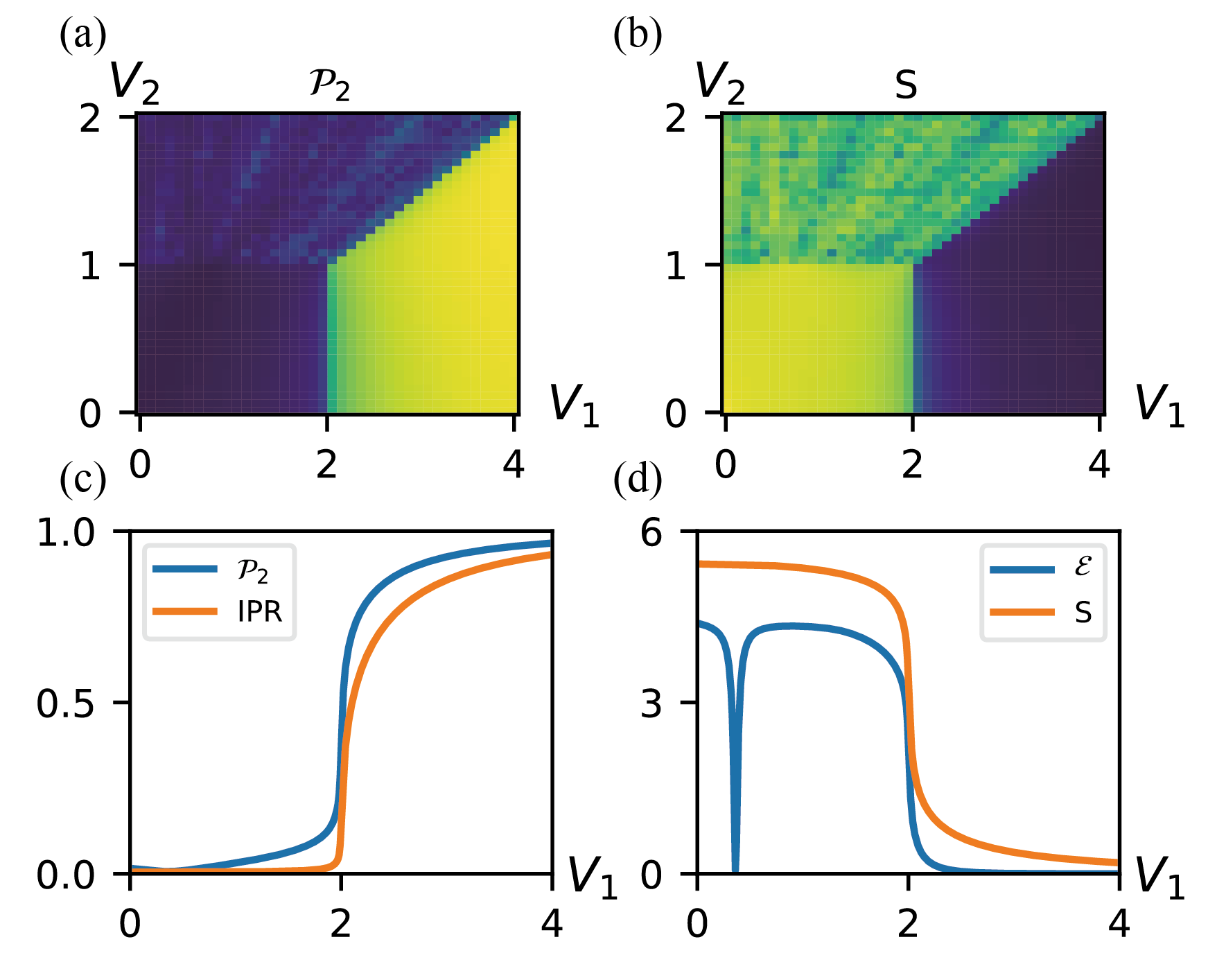}
    \caption{Phase diagrams of the generalized AAH model obtained using the persistent homology feature norms $\mathcal{P}_2$ (a) and Shannon entropy $S$ (b) of the ground state. (c,d) Comparison between $\mathcal{P}_2$ and IPR (c) and betweeen the persistent entropy $\mathcal{E}$ and Shannon entropy $S$ along a 1D cut of the phase diagram. Only $\mathcal{E}$ detects the transition to order at $V_1 = 2 V_2 \cos (Q/2)$.}
    \label{fig:phases_2}
\end{figure}

Figure~\ref{fig:phases_2}(a,b) shows the phase diagrams obtained using feature lifetime norm $\mathcal{P}_2$ [Eq.~\eqref{eq:pnorm}] and the ground state's Shannon entropy [Eq.~\eqref{eq:shannon}]. $\mathcal{P}_2$ is poor at distinguishing the extended and critical phases, similar to IPR. On the other hand, $S$ provides similar contrast to the persistent entropy, but does not detect the ordered line. This suggests that the main advantage of the persistent homology approach may be in distinguishing order from disorder. We illustrate this further by plotting all four eigenstate measures along 1D cuts of the full phase diagram in Fig.~\ref{fig:phases_2}(c,d). While all measures can reproduce the correct phase boundaries, the persistent entropy provides stronger contrast between the critical and extended phases and can also distinguish order from disorder in the ground state, visible as a prominent dip in $\mathcal{E}$ in Fig.~\ref{fig:phases_2}(d).

\begin{figure}
    \centering
    \includegraphics[width=\columnwidth]{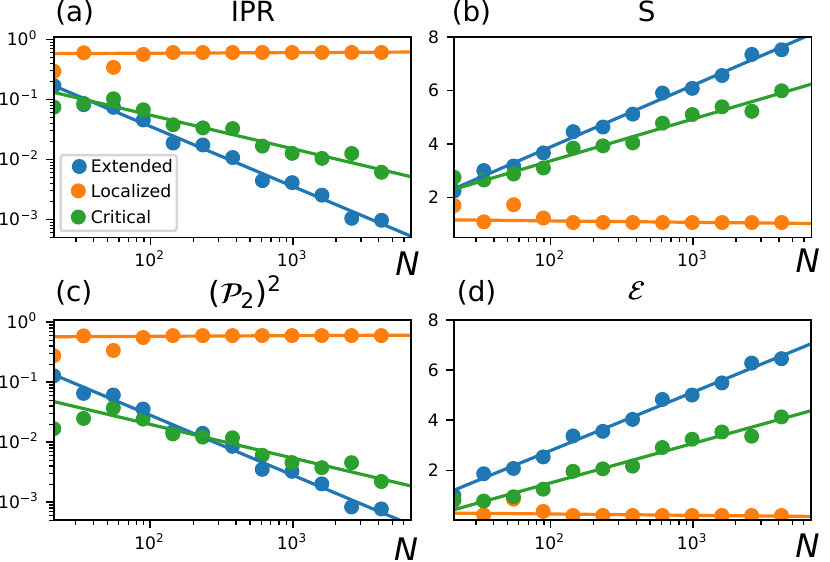}
    \caption{Finite size scaling of the lowest energy eigenstate's IPR (a), Shannon entropy $S$ (b), $(\mathcal{P}_2)^2$ (c), and persistent entropy $\mathcal{E}$ (d) in the three phases. The standard measures (a,b) and persistent homology-based measures (c,d) exhibit the similar scaling feature.}

    \label{fig:fss}
\end{figure}

For a given eigenstate measure the contrast between the different phases will depend on the system size $N$. For example, the generalized inverse participation ratios will scale in the large $N$ limit as $\mathrm{IPR}_q \propto N^{d_q(1-q)}$, where $d_q$ is the fractal dimension of the $q$th eigenstate moment~\cite{Anderson_review,Roemer2011}. Thus, $d_q$ can be obtained by studying the scaling of an eigenstate measure with $N$. Fig.~\ref{fig:fss} plots the finite size scaling of the measures of the lowest energy eigenstate, taking the system size $N$ to be Fibonacci numbers and corresponding rational approximants of $Q$ to minimize finite size effects. We observe that $\mathrm{IPR}$ and $(\mathcal{P}_2)^2$ and $S$ and $\mathcal{E}$ exhibit identical scaling exponents, up to the accuracy of the obtained power law fits.

Both the standard and persistent homology-based measures give similarly faithful estimates for the fractal dimensions of the eigenstates by taking the large $N$ limit. On other hand, based on the offsets between the fits obtained for the different phases, it appears that, for fixed $N$, the $\mathrm{IPR}$ (persistent entropy) performs better at distinguishing the extended and critical phases than P2 (Shannon entropy). Since both approaches yield the same fractal dimensions $d_{1,2}$ we anticipate their multifractal spectra will also coincide. A more detailed study of multifractal properties probed using persistent homology is an interesting direction for future research.

\section{Propagation dynamics}
\label{sec:dynamics}

\begin{figure}
    \centering
    \includegraphics[width=\columnwidth]{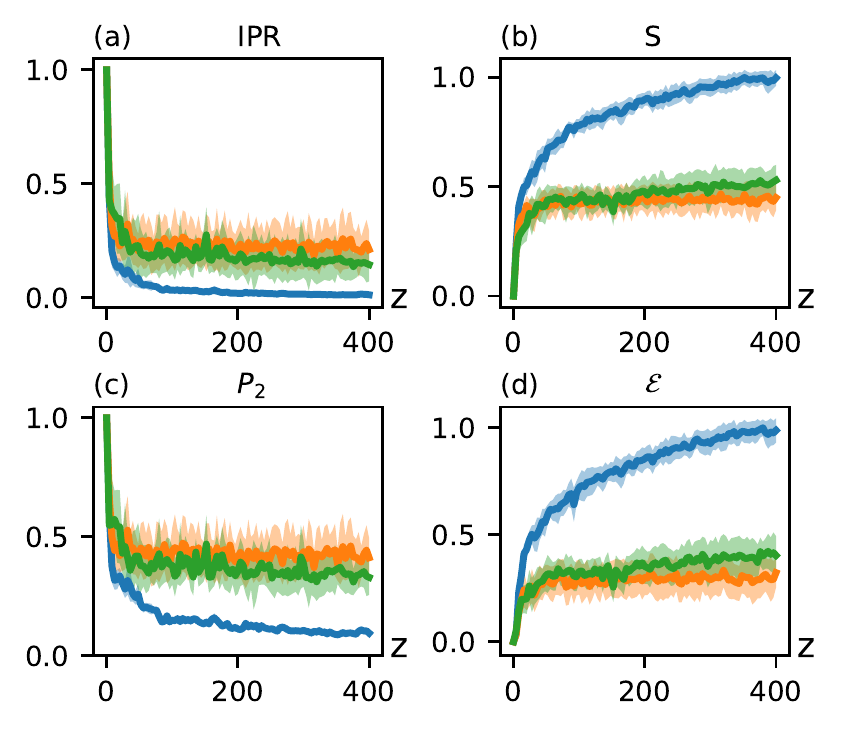}
    \caption{IPR (a), Shannon entropy (b), $\mathcal{P}_2$ (c), and persistent entropy $\mathcal{E}$ (d) as a function of the propagation distance $z$ in the three phases (blue - extended [$V_1=1.8$, $V_2=0.5$]; orange - localized [$V_1=2.2$, $V_2 = 0.5$]; and green - critical [$V_1=1.0$, $V_2=1.5$]). Solid lines indicate mean values over an ensemble of $k$ values, shaded regions one standard deviation.}
    \label{fig6}
\end{figure}

Next we study how persistent homology analysis can be used to characterize the propagation dynamics. For this, we consider the evolution of a localized (single waveguide) excitation, described by Schr\"odinger equation $i \partial_z \psi_n(z) = \hat{H}\psi_n(z)$, with $\psi_n(0) = \delta_{n,N/2}$, where $N$ is the total number of lattice sites. We compute IPR, $S$, $\mathcal{P}_2$, and $\mathcal{E}$ of the $z$-dependent intensity profiles $|\psi_n(z)|^2$ in Fig.~\ref{fig6}. To obtain statistics and uncertainty estimates, we construct an ensemble by varying the parameter $k$ appearing in Eq.~\eqref{eq:aah}, which is equivalent to spatial translations of the input beam. 

The standard approach towards distinguishing the phases of $\hat{H}$ using the propagation dynamics is to consider the scaling of IPR, i.e. using a power law fit $\mathrm{IPR} \propto z^{-\alpha}$, where the fitted exponent $\alpha$ can distinguish localized, diffusive, and ballistic transport regimes~\cite{loc_review,AAH_experiment}. In the localized phase there is an initial transient expansion until the wavepacket width is comparable to the localization length $\xi$, after which spreading stops, corresponding to $\mathrm{IPR} \sim 1/\xi$ and $\alpha = 0$. In delocalized phase a single site excitation will spread indefinitely, corresponding to ballistic expansion with $\mathrm{IPR} \propto z$, i.e. $\alpha = 1$. In the critical phase, the transport is approximately diffusive, with $\alpha \approx 1/2$\cite{Longhi2021}.

Consistent with these expectations, in the localized phase in Fig.~\ref{fig6} (orange curves) the IPR does not decay to zero but saturates at a finite value, indicating localization. Similar saturation is seen in $\mathcal{P}_2$. Meanwhile, $S$ and $\mathcal{E}$ grow the fastest in the extended phase and saturate at a low values in the localized phase. However, it is noticeable that there is a subtle difference between Fig.~\ref{fig6}(a,b) and Fig.~\ref{fig6}(c,d) after a long propagation distance: the persistent homology measures show a quantitatively larger difference between the localized phase and the extended and critical phases.

\begin{figure}
    \centering
    \includegraphics[width=\columnwidth]{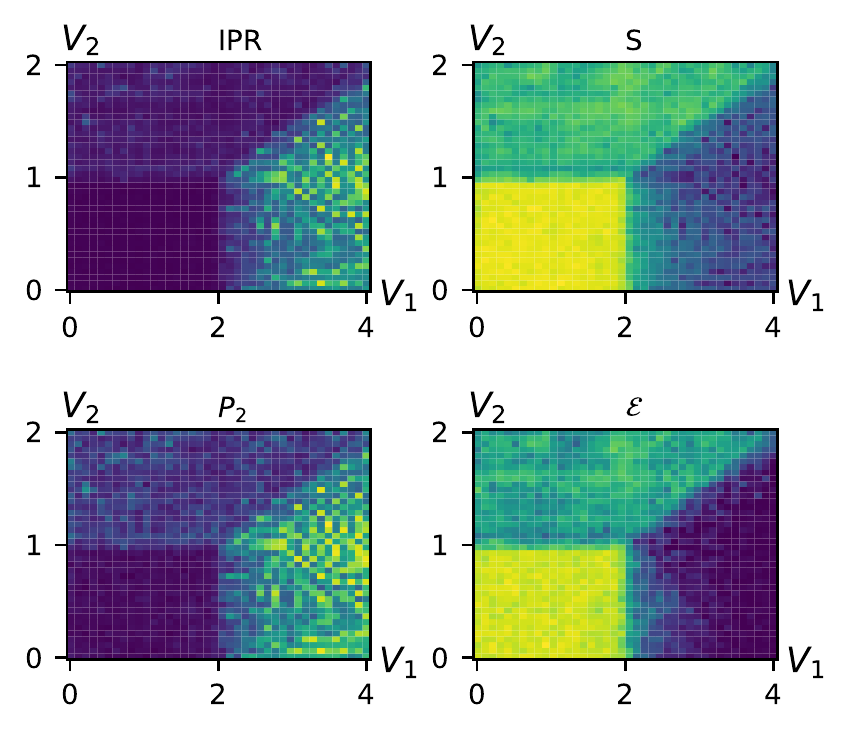}
    \caption{Phase diagrams obtained from the beam propagation simulations, using the output intensity profile $|\psi_n(L)|^2$. Propagation distance $L$ = 300.}
    \label{fig4}
\end{figure}

We repeat this procedure, scanning over the model parameters $V_1$ and $V_2$ and computing IPR, $S$, $\mathcal{P}_2$ and $\mathcal{E}$ after the wavepacket has propagated a sufficiently large distance. Fig.~\ref{fig4} illustrates the phase diagrams reconstructed from the beam propagation simulations. We see that both the conventional measures (IPR, $S$) and the persistent homology measures ($\mathcal{P}_2$, $\mathcal{E}$) are able to distinguish the dynamics in the three phases. The most striking difference compared to Fig.~\ref{fig1} is that the ordered line $V_1 = 2 V_2 \cos (Q/2)$ is not detected by the propagation dynamics. The reason for this is that along this line, the quasiperiodic potential is only transparent to the low energy modes; the single site initial state excites a superposition of all the modes. Thus, the field becomes disordered after propagating a sufficiently long distance.

\begin{figure}
    \centering
    \includegraphics[width=\columnwidth]{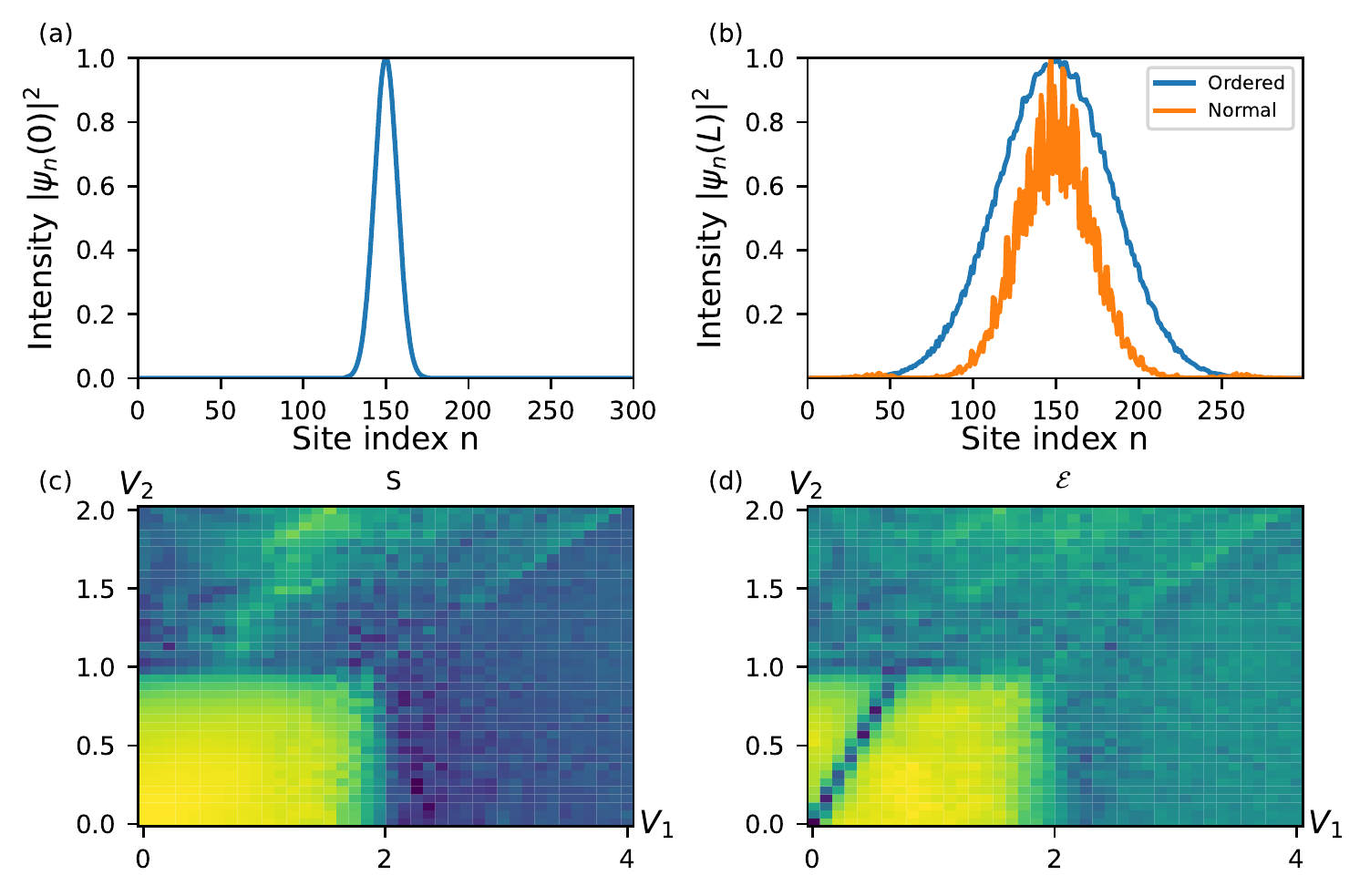}
    \caption{(a) Amplitude profile of low energy wavepacket initial state and (b) output field profiles after propagating a distance $L=250$ on the ordered line (blue; $V_1 = 0.3, V_2 
\approx 0.41$) and elsewhere in the extended phase ($V_1 = 0.3, V_2 = 0.8$). (c,d) Phase diagrams obtained from the output beam profile's Shannon entropy $S$ (c) and persistent entropy $\mathcal{E}$ (d).}
    \label{fig:wave}
\end{figure}

To determine whether the ordered line can be feasibly observed in experiment we construct a finite size Gaussian beam with width $w=10$ and alternating phase between neighboring sites, $\psi_n(0) = (-1)^n \exp[ - (n-N/2)^2 / (2 w^2) ] $, shown in Fig.~\ref{fig:wave}(a). Owing to the similarity between this profile and the ground state, in the extended phase this initial state mainly excites the low energy modes. Fig.~\ref{fig:wave}(b) shows two intensity profiles obtained after propagation at ordered and disordered points in the extended phase. Thanks to the suppression of the high energy modes there is a clear difference in the output intensity profiles; the profile is smooth on the ordered line, whereas elsewhere in the extended phase rapid intensity fluctuations are generated. The Shannon entropy of the output intensity profiles shown in Fig.~\ref{fig:wave}(c) does not distinguish these two profiles, whereas the persistent entropy in Fig.~\ref{fig:wave}(d) correctly identifies the ordered line, showing that the persistent homology method may also be useful for distinguishing ordered from disordered propagation dynamics without requiring ensemble averages.

\section{Conclusion}
\label{sec:conclusion}

We have studied how the persistent homology analysis can be applied to characterize the eigenstates and propagation dynamics of tight binding models. Focusing on the example of the generalized Aubry-Andr\'{e}-Harper model, we have shown how point summaries of the persistence diagrams can be used to distinguish its three phases (extended, localized, critical), as well as distinguish ordered from disordered regimes. This is despite the fact that the persistence diagrams form a highly compressed representation of the eigenstates, i.e. containing only information about their local maxima and minima. Persistent homology-based observables such as the persistent entropy may be a useful tool for detecting novel phases and parameter regimes of complex lattice models.

We focused on the case of a one-dimensional lattice for ease of understanding. However, persistent homology can also be readily applied to higher dimensional intensity profiles, where it can not only characterize the number of high intensity clusters, but also detect higher-dimensional topological features such as holes and loops. This could be useful for distinguishing localized, extended, and compact localized modes in disordered flat band lattices such as the Lieb lattice and its higher dimensional generalizations~\cite{Lieb2D,Lieb3D}.

Another interesting direction would be to investigate whether persistent homology-related observables are amenable to analytical or semi-analytical calculations. For example, in 1D tight binding lattices the energy-dependent localization length can be computed without diagonalizing the Hamiltonian by evaluating a recurrence relation satisfied by the eigenstates~\cite{loc_review,Anderson_review}. Analytical results may reveal to which specific features of eigenstates persistent homology is most sensitive.

Since persistent homology captures multiscale properties of wavefunctions, the techniques developed here may be useful for characterizing the response of active photonic systems, such as supercontinuum and frequency comb generation in microring resonators, where the intensity can be measured over a huge range of scales~\cite{comb}. Thus, we envision the techniques could be further developed for advanced studies, which may find applications in deep-learning-based photonic design and AI-empowered nanophotonics that will continue to grow in the coming decade~\cite{final1,final2}.

\section*{Acknowledgements}

This research was supported by the National Natural Science Foundation of China (12134006), National Research Foundation, Prime Minister's Office, Singapore, the Ministry of Education, Singapore under the Research Centres of Excellence Programme, China Postdoctoral Science Foundation (BX2021134, 2021M701790) and the Polisimulator project co-financed by Greece and the EU Regional Development Fund.

\end{document}